# Hybrid Graphene/Carbon Nanofiber Wax Emulsion for Paper-based Electronics and Thermal Management

*Xinhui Wu, Pietro Steiner, Thomas Raine, Gergo Pinter, Andrey Kretinin, Coskun Kocabas, Mark Bissett\*, Pietro Cataldi\**

X Wu, P. Steiner, T. Raine, G. Pinter, A. Kretinin, C. Kocabas, M. Bissett, P. Cataldi
Department of Materials, National Graphene Institute, University of Manchester, Oxford Road, Manchester, M13 9PL UK
E-mail: pietro.cataldi@manchester.ac.uk; mark.bissett@manchester.ac.uk

Materials for electronics that function as electrical and/or thermal conductors are often rigid, expensive, difficult to be sourced and sometimes toxic. An electrically and thermally conductive nanocomposite that is lightweight, flexible and eco-friendly could improve the environmental friendliness of the electronics sector and enable new applications. Considering this, we have fabricated electrically and thermally conductive flexible materials by functionalizing paper with nanocarbon conductive inks. Carnauba wax is emulsified in isopropyl alcohol and mixed with graphene nanoplatelets (GNPs) or with hybrids of GNPs and carbon nanofibers (CNFs). The percolation threshold of the hybrid samples is lowered compared with the pure GNPs composites, due to their increased filler aspect ratio. The hybrid samples also exhibit superior bending and folding stability. Densification of the coating to decrease their sheet resistance enables them to achieve as low as ~ 50 Ω sq$^{-1}$ for the GNP-based paper. The densification procedure improves the bending stability, the abrasion resistance, and the electromagnetic interference shielding of the paper-based conductors. Finally, the compressed samples show an impressive enhancement of their thermal diffusivity. The flexible and multifunctional nanocarbon coated paper is a promising electronic conductor and thermally dissipative material and, at the same time, can increase the environmental sustainability of the electronics sector.

# 1. Introduction

The omnipresence of electronics in daily life is unquestionable. Electronic devices improve the quality of life and a world without electrically powered infrastructure, smartphones and laptops is unimaginable. The thriving electronics sector will be propelled soon by the transition to conformable/flexible electronics and to wearable integrated technologies that will build a more intimate relationship between the electrical components and human beings/bodies.[1, 2] On the other hand, the ubiquity of electronic devices constitutes a problem considering their end of life and potential environmental impact of disposal.[3, 4] Indeed, electronic waste is growing at the fastest rate and, under a worst-case scenario, is forecast to reach a production of 120 million metric tons per year in 2050.[5, 6] Furthermore, materials used for electronics are often expensive and/or difficult to be sourced and sometimes are toxic to humans and the environment.[1, 2, 7-9] To facilitate the transition towards truly portable/conformable devices, tackling at the same time the ever-increasing problems of electronic waste management and the unsustainable nature of the electronics sector, flexible, cheap, abundant and environmentally friendly material for electronics should be designed.[10, 11]

Although the fields of flexible electronics have reached substantial breakthroughs recently, combining flexibility and environmental friendliness to realize next-generation materials for electronics is still an ongoing challenge.[10] To address this, one of the most promising materials is simple paper, which is constituted of cellulose, the most abundant biodegradable biopolymer obtained from renewable and sustainable sources.[2, 4, 12-18] Paper can be functionalized with conductive materials in the form of inks that impart their electrical conductivity to the substrate while maintaining its flexibility.[19, 20] Metal-based nanomaterials [21-23], conductive polymers [24-26], liquid metals [27, 28] and carbon nanoparticles [24, 29-32] are among the conductors most apt to realize conductive inks for paper functionalization. Nevertheless, drawbacks are still present. For example, metal-based nanomaterials are expensive and their dispersion in liquids is laborious.[29, 33] The adhesion of the inks on the paper is often guaranteed using long-lasting persistent materials that decrease the environmental friendliness and the electrical conductivity of the flexible conductor.[15, 33, 34] Moreover, the liquids employed for these inks are often toxic considering the green solvent categorization.[5, 33-35] Finally, the functionalization of the paper is sometimes achieved through complicated procedures that hinder the industrialization of these flexible conductors.[14, 15, 36] Taking this into account, there is still margin for improvement. Emulsification is emerging as a convenient method to fabricate inks with applications as diverse as batteries [37], water repellent textiles [38], conductive materials [39-42], pharmaceutical products [43, 44], tissue engineering [45], superhydrophobic coatings [39, 42, 46, 47], agricultural additives [48, 49] and cosmetic goods [50]. Wax-based emulsions are advantageous due to their simple fabrication and versatility.[51-55] Natural waxes such as beeswax and carnauba are attractive for their biodegradability, large scale availability and waterproof features.[55, 56] In particular, carnauba wax (CW) shows superior mechanical properties and a higher melting temperature compared with beeswax (82–86 °C vs 62–64 °C). CW was previously used to improve the gas/moisture barrier properties in the food packaging industry [51, 56-59], to create edible coatings that can increase the shelf-life of food [60, 61] and to manufacture superhydrophobic materials [62-64].

Conductive inks made with carbon nanomaterials are a convenient alternative to metal-based inks, considering, for example, that silver is extremely expensive and that copper can easily oxidize and lose its electrical conductivity.[2, 29, 65] Among carbon nanomaterials, graphene nanoplatelets (GNPs) are an ideal candidate for the creation of industrially available inks in light of their large volume of production, low cost, and versatility.[2] Furthermore, GNPs, as opposed to graphene oxide, does not require any thermal annealing procedure to achieve satisfactory electrical conductivity. This is a crucial requirement for paper-based electronics since a high temperature can degrade the functionalized substrate.[34] Recently, high aspect ratio carbon nanofibers (CNFs) also showed their potential in the creation of conductive inks.[24, 30, 42, 66]

Another advantage that GNPs and CNFs have in the formulation of conductive inks is that their employment can facilitate the creation of multifunctional materials. They possess a diverse selection of properties that include high electrical and thermal conductivities, remarkable mechanical features and flexibility, to cite just a few.[2, 30, 42, 66] Considering this, and their low cost compared to metal-based nanomaterials, GNP- and/or CNF-based nanocomposites could substitute gold and silver as electrodes and/or thermal dissipators in the electronics sector and, coupled with flexible materials, they could be used as flexible electrical/thermal conductors.

Here, we report the fabrication of electrically and thermally conductive flexible materials by simply functionalizing paper with conductive inks. CW was mixed with pure GNPs or with hybrids of GNPs and CNFs. The inks were produced through the emulsification of CW in isopropyl alcohol. After the conductive nanoparticle inclusion, the inks were spray-coated on the paper substrates and heated to melt the CW and stabilize the flexible conductor. The percolation threshold of the hybrid samples was lowered of 4 weight % compared with the pure GNPs-based samples. Both samples showed remarkable bending stability, with the hybrids showing an increase in the sheet resistance of less than 10% after 30000 bending cycles. The hybrid samples showed better preservation of their electrical conductivity compared to GNPs-based samples after repeated folding-unfolding events. Densification of the conductive layer after spray coating improved the electrical conductivity of the flexible conductor. The lowest value of sheet resistance was achieved by the GNP-based paper that reached 50 $\Omega$ sq$^{-1}$ after pressing. The pressing procedure also improved the bending stability of the samples and, importantly, enhanced their abrasion resistance. Furthermore, it improved also the electromagnetic interference (EMI) shielding of the materials. The compressed samples also displayed an impressive 200% enhancement of their thermal diffusivity, a remarkable improvement compared with the low thermal diffusivity of the paper substrate. The obtained multifunctional paper-based conductors could find applications in flexible electronics and thermal dissipation and, at the same time, increase the environmental sustainability of the electronics sector.

## 2. Results and Discussion

A schematic of the preparation of the conductive ink is presented in Figure 1a. The emulsions were fabricated in isopropyl alcohol due to its environmental friendliness compared with other organic solvents.[35] To emulsify the CW and create a wax in alcohol emulsion, the polymer was melted heating up the vial to ~ 85 °C, above the melting temperature of the wax as confirmed by DSC analysis (Figure S1). At this temperature, the emulsion was transparent and yellowish. A tip sonication process was briefly performed to homogeneously disperse the wax droplets in the alcohol. A similar procedure is already documented to produce emulsified waxes in alcohol.[67] When cooled down, the dispersion became white, indicating that light was scattered equally by the micro-particles of CW that were formed after cooling. These particles exhibited dimensions mostly below 10 μm$^2$ which corresponds to a lateral size lower than 4 μm (See Figure S2). The addition of GNPs or hybrids of GNPs and CNFs (ratio 1:1) and a tip sonication process was sufficient to create a conductive ink. These bi/tri-component dispersions were spray-coated on top of printing paper and afterward an additional heat gun process at approximately 100 °C was carried out to melt the wax. Such temperatures were not damaging to the samples, as confirmed by TGA analysis (See Figure S3). This additional heating treatment also stabilized the coating adhesion on the paper. Indeed, without this treatments samples were found to be unstable and difficult to handle.

From the SEM analysis the morphology of the pure paper substrates shows 10-100 μm thick cellulose fibers filled with sizing agents (typically agglomerates of starch observable in Figure 1b) that enhances the printability and water-resistance of the fibrous network.[68, 69] After spraying of the GNPs and hybrid inks, the paper surface appears to be carpeted with randomly oriented GNPs or GNPs-CNFs webs, as shown in Figure 1c and 1d, respectively. This surface texturing is typical of GNPs-based spray-coated dispersions on paper.[16, 68] Nanofibers bridge the graphene flakes one with the others determining a morphological characteristic that is important for the electrical properties and deformation stability of the samples (see Figure 2 for more details). The cross-section of the pure paper appears as a disordered fibrous mat while the coating, thanks to the melting of the CW during heating, appears as a more compact layer after the cut. As can be seen in the inset of Figures 1f and 1g, the thicknesses of the inks vary between 10 to 20 microns due to the roughness induced by the presence of the randomly oriented GNPs.

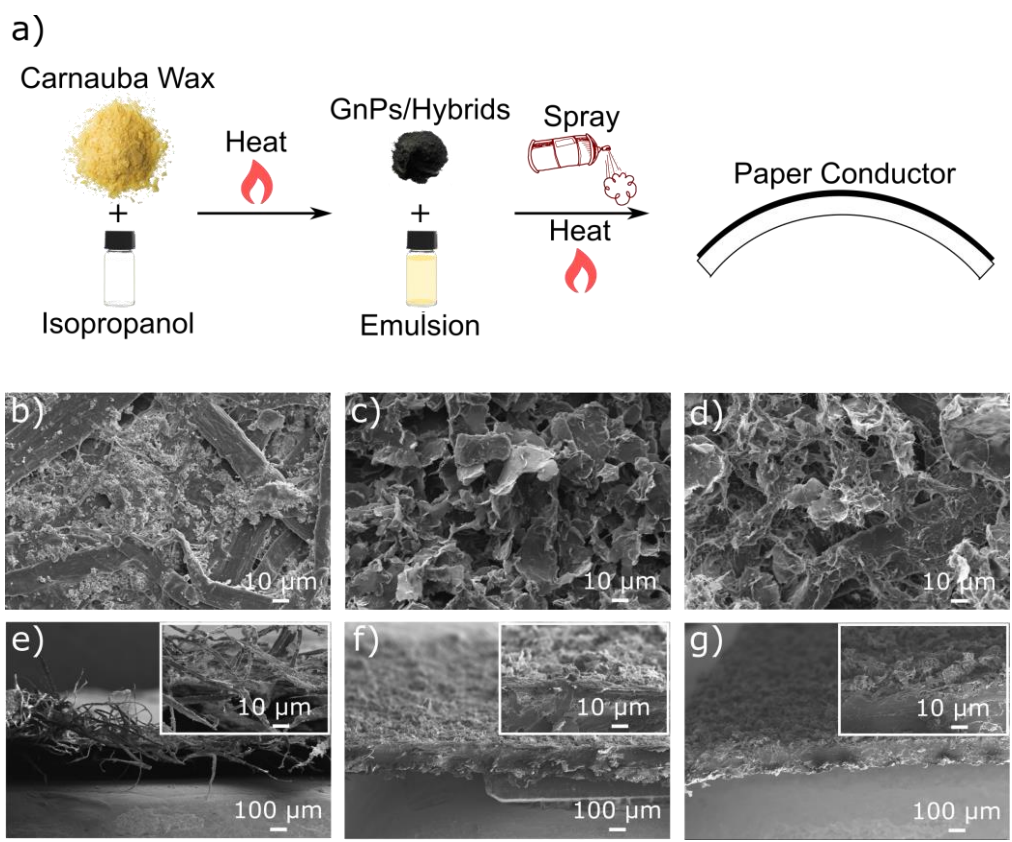

**Figure 1**. a) Schematic for the preparation of the conductive ink. b) - d) SEM morphologies of the pure paper, the GNP based material and of the hybrid, respectively. e) - g) cross-section of the samples presented in b) – d).

The electrical percolation threshold determines the minimum amount of nanofillers needed to achieve electrical conductivity.[70-72] Figure 2a plots the sheet resistance of the conductive coating with increasing loading (wt.%) of nanofillers relative to CW mass. For the pure GNP containing coating, the sheet resistance was above $10^{10}$ Ω sq$^{-1}$ up to 4 wt.% concentration, while it dropped six orders of magnitude between 6 and 8 wt.% indicating that the percolation threshold was at this loading. The sheet resistance further diminished with increasing loadings, reaching values of ~ 300 Ω sq$^{-1}$ at 40 wt.% GNPs. On the other hand, the hybrid samples displayed a reduced percolation threshold compared with the coating containing only GNPs. There is an abrupt decrease of several orders of magnitude at 2 wt.% nanofiller. Increasing the amount of nanofillers led to a further decrease in the sheet resistance, reaching ~ 800 Ω sq$^{-1}$ at 40 wt.% concentration. To analyze the percolation behavior of the conductive coating, we used percolation theory.[68, 70] Considering the nanofillers as randomly distributed, it was possible to use the following equation to describe the percolation threshold:

$$\sigma = \sigma_0 (\phi - \phi_c)^t, \qquad (1)$$

Where σ is the electrical conductivity, ϕ is the nanofiller concentration, $\phi_c$ is the percolation threshold and t is the universal critical exponent.[73] Employing 8 wt.% and 2 wt.% as $\phi_c$ for the pure GNP and Hyb case, respectively, we plotted the log-log of σ as a function of $\phi - \phi_c$ and performed a linear fit (Figure S4), from this we extracted the universal critical exponent. For both coatings, t ≈ 1.3. The universal critical exponent is associated with the micro and nano assembly of the percolated systems and the calculated value indicates the construction of 2D networks of nanofillers in the conductive coating applied to the paper.[70, 74]

The characterization of the electrical properties of the paper-based conductors against mechanical deformation is critical for effective implementation of flexible electronics. Therefore, the electrical features of the conductive paper were measured before and after bending (Figure 2b) and folding (Figure 2c). These tests were performed on the specimens with 30 wt.% of nanofillers relative to CW (referred from now on as G and Hyb for the pure

GNPs and hybrid samples, respectively). 30 wt.% samples were chosen since lower loadings of nanofillers displayed slightly worse deformation stability (see Figure S5) and considerably lower electrical conductivity (Figure 2a). In addition, higher loadings of nanofillers were not chosen as they did not show substantially improved electrical properties compared to 30 wt.% and were found to clog the spray nozzle during fabrication. It is worth noting that all the flexible conductors showed paper-like mechanical properties (see Figure S6) and therefore were appealing for applications in flexible electronics and for the electro-mechanical test mentioned above.

The bending tests were performed by measuring the electrical surface resistance (R) in a flat configuration, and when the paper conductor was bent at 0.4 cm bending radius. During the first 150 bending cycles, the ratio $R/R_0$, increased by ~50 % for the G sample in both the flat and curved configuration. After 30000 bending cycles, $R/R_0$ increased by 60 and 70% in the flat and curved case, respectively. Conversely, the Hyb samples exhibited a lower change after 150 bending cycles, displaying a value of $R/R_0$ which increased by only of the 7 and 9 % in the flat and curved configuration, respectively. After 30000 bending cycles, the normalized resistance increased by only the 8 and 10% in the flat and curved configuration, respectively. The enhanced bending stability of the Hyb samples is a result of the CNFs 1D nanostructure that facilitates the connection between GNPs even after repeated bending cycles and in a curved configuration. Their exceptionally high aspect ratio is able to bridge the GNPs, thus maintaining highly conductive nanofiller networks. Considering practical applications, it is relevant that a conductive paper can sustain tens of thousands of bending cycles without any marked drop of electrical conductivity.

Folding stability is also a critical requirement for paper-based electronics. Therefore, cyclic 180° folding–unfolding tests were performed on the samples.[68, 71] At each cycle, the folded edge was pressed with a 1.5 kg weight to enhance the harshness of the test.[68, 71] The ratio $R/R_0$ perpendicular to the fold line for 20 fold–unfold cycles is presented in Figure 2c. After the first two cycles, the electrical resistance of the G samples increased by roughly two orders of magnitude. From the third cycle, the G specimens were not conductive anymore. On the other hand, the Hyb sample showed a significantly better folding stability compared with the pure GNP specimen, reaching a two order of magnitude increase of the ratio $R/R_0$ only after 15 folding cycles. The fold line was macroscopically evident due to the formation of cracks that were responsible for the resistance increase. However, the percolating network of the hybrid specimen was better preserved after fold due to the high aspect ratio of the nanofibers and to the conductive bridge formed between the GNPs.

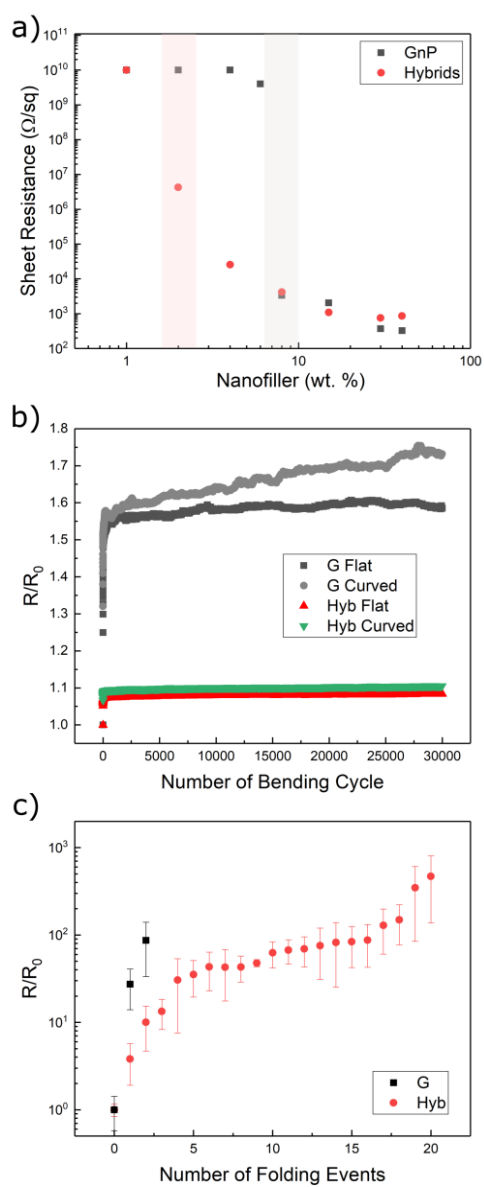

**Figure 2**. a) Sheet resistance of the conductive paper as a function of the nanofillers loading. The red and grey areas indicate the percolation region for the hybrids and pure GNP coating, respectively. b) and c) show the bending and the folding stability of the electrical resistance, respectively.

Densification procedures are known to be beneficial for the electrical conductivity of graphene-based materials/coatings.[68, 75, 76] Therefore, we investigated the effect of applied pressure on the paper coating. After spraying, the samples were compressed and subsequently treated with a heat gun to melt the CW and bind the substrate and the nanoparticles (see materials section for more details). Pressing our paper-based conductors led to a flattening of the conductive coating. The SEM morphologies displayed a reduced roughness and an improved homogeneity of the disposition of the nanofillers on top of the paper substrate, as shown in figure 3a and figure S7. Such reduced disorder boosted the direct contact between the nanofillers and created an efficiently interlinked micro-network of conductors. Consequently, the path needed for the electrons to transit from point to point of the coating was decreased. This morphological change was reflected also in the decrease of the sheet resistance of the coating as a function of the applied pressure as displayed in figure 3b. The G sample decreased its sheet resistance linearly (~2.6 ($\Omega$ sq$^{-1}$)/bar) passing from ~ 400 $\Omega$ sq$^{-1}$ with no pressure applied to ~ 50 $\Omega$ sq$^{-1}$ after a press at 112.5 bar. On the other hand, the Hyb specimens exhibited a pronounced jump of ~ 600 $\Omega$ sq$^{-1}$ with the first pressure step, passing from ~ 800 $\Omega$ sq$^{-1}$ to ~ 200 $\Omega$ sq$^{-1}$. Further increasing the pressure to 112.5 bar led only to a slight decrease of the sheet resistance that reached ~ 120 $\Omega$ sq$^{-1}$. Since the best electrical conductivities were reached at the highest pressure, the pressing procedure was subsequently kept at 112.5 bar. It is worth noting that the infiltration of the conductive nanofiller layer inside the cellulose fibers was prevented due to the sizing agents present on top of the paper, as shown in Figure 1b and in the cross-section image of the pressed samples (Figure S8). The pressing procedure was also beneficial for the bending stability of the paper-based conductors (Figure S9). After pressing, the G sample had increased normalized electrical surface resistance by only 8 and 40% after 30000 bending cycles in the flat and curved configuration, respectively. With pressing, the Hyb sample maintained essentially unaltered bending stability, increasing its normalized electrical surface resistance by about 10% for both the flat and curved state after 30000 bending cycles.

The most common abrasion-resistant materials are realized with hard coatings.[77] Flexible and lightweight materials that maintain their electrical conductivity after repeated abrasion cycles are still a great challenge to be manufactured, yet could satisfy unmet requirements.[71] Therefore, repeated abrasion cycles were performed on the samples, as shown in Figure 3c. A comparison before and after pressing was also performed. The G sample after one abrasion cycle increased its resistance by one order of magnitude and was not electrically conductive following another abrasion cycle. The Hyb sample increased its electrical resistance by four orders of magnitude during the first three abrasion cycles and lost its electrical conductivity afterward. The loss of electrical conductivity was driven by the removal of the conductive coating, as shown in Figure S10. In contrast, the G and Hyb samples after the press (named G-P and Hyb-P, respectively) maintained their electrical properties as the resistance remained essentially unchanged during the first 40 abrasion cycles. From the $40^{th}$ to the $160^{th}$ cycle their sheet resistance was increased by one and three orders of magnitude, respectively. The remarkable improvement of the abrasion resistance of the pressed samples is certainly associated with the low friction coefficient of flat surfaces compared with un-pressed, rough samples. Moreover, the better performance of the G samples compared with the Hyb samples is related to the high tribological performance that such nanomaterials provide.[78, 79]

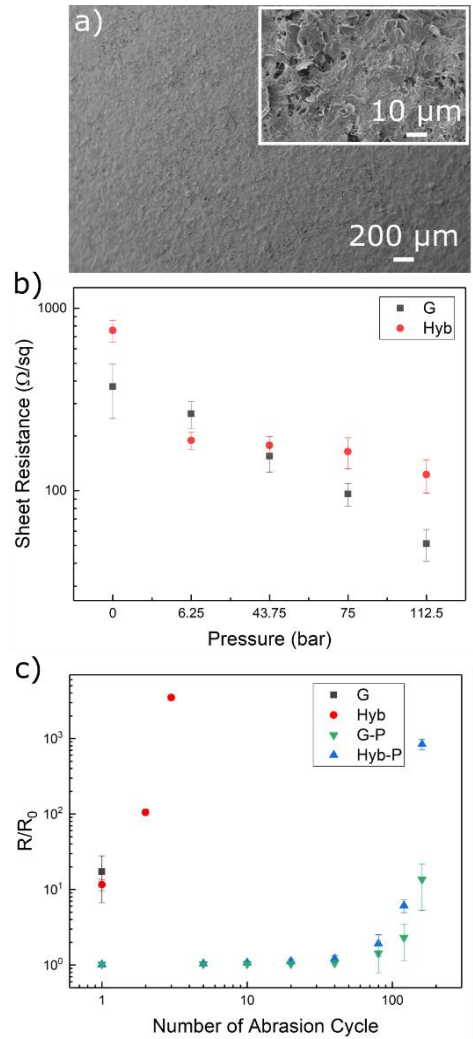

**Figure 3**. a) SEM images of the top view of the Hyb sample after pressing at 112.5 bar. b) Sheet resistance variation as a function of the pressure applied. c) Abrasion resistance of the electrical resistance of the sample before and after the press.

Materials that exhibit multifunctional properties are of particular interest because they can be employed for myriad applications, providing a great advantage for both manufacturers and consumers. Recently Kargar et al. demonstrated a bi-functional graphene composite that has outstanding electromagnetic shielding performance and high thermal conduction properties.[80] It consisted of rigid epoxy resin mixed with few-layered graphene. An electrically and thermally conductive composite that is lightweight, flexible and eco-friendly could further expand the number of possible applications of this class of material.[80-82] Considering this, we measured the EMI shielding of the conductive coated papers between 8 and 12 GHz (see Figure S11) and we discovered that the pressing procedure increased the shielding efficacy of both the Hyb and G samples, reaching an EMI shielding of ~ 8 dB in the best case. Previously, a pressing procedure was already found to be helpful for EMI shielding due to the induced alignment of carbon-based fillers.[76] In particular, after pressing, the G sample displayed improved EMI shielding of approximately the 40% while the pressed Hyb sample improved by 60%. We tested also the in-plane thermal diffusivity (TD) of the G and Hyb samples, as shown in Figure 4. The neat paper showed a TD of 0.5 mm$^2$/s. The G and Hyb samples exhibited a TD that was of ~ 0.6-0.7 mm$^2$/s, only slightly higher than the bare substrate. A large difference was recorded after the densification, highlighting that it was this procedure that was boosting the thermal conduction properties and not the nanoparticle composition within the coating. Indeed, G-P and Hyb-P samples reached the same values of 1.4 - 1.5 mm$^2$/s, within experimental error, an increase of the 200% compared with the neat paper. After the pressing procedure, the thermal diffusivity is greatly enhanced due to the increasing number of connections between the nanofillers, thus creating thermally conductive pathways inside the coating.

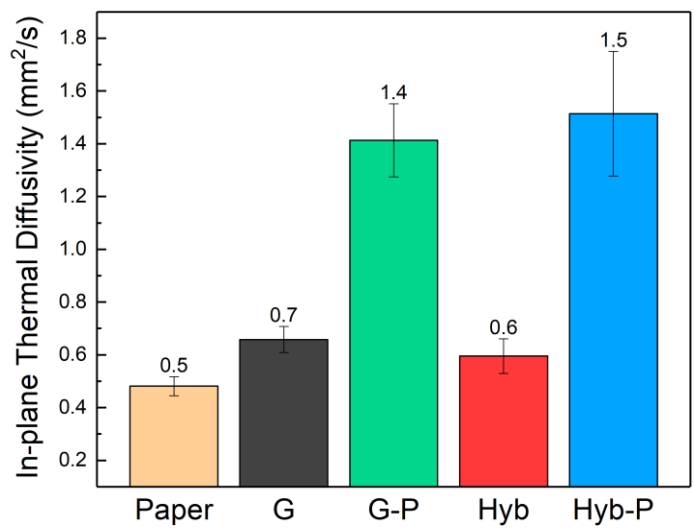

**Figure 4**. Thermal diffusivity of the pure substrate (paper) and of the GnP- and hybrid-based materials before (named G and Hyb) and after pressing (G-P and Hyb-P).

## 3. Conclusions

In pursuit of realizing a flexible electrical and thermal conductor, we functionalized paper by coating with conductive inks obtained through emulsifying carnauba wax with GNPs or with hybrids of GNPs and CNFs in isopropyl alcohol. After subsequent heat treatment, the carnauba wax acted as a binder between the carbon nanomaterials and the paper substrate. The percolation threshold of the hybrid samples decreased by 4 wt.% compared with the pure GNPs-based samples, percolating at 2 wt.% hybrid nanofiller loading. Both samples showed remarkable bending stability, with the hybrids displaying an increase of the sheet resistance of less than 10% after 30000 bending cycles. After repeated folding-unfolding events, the hybrid sample preserved its electrical conductivity better when compared with the pure GNPs samples. Pressing the conductive paper decreased the sheet resistance of the flexible conductors. The lower sheet resistance was shown by the bare GNP-functionalized paper that reached 50 $\Omega$ sq$^{-1}$ after the press. The pressing procedure improved the bending stability of the samples and importantly enhanced also their abrasion resistance. Furthermore, pressing improved the electromagnetic interference shielding of the material. Finally, the compressed samples showed an impressive enhancement of their thermal diffusivity of 200%. The obtained flexible and multifunctional paper-based conductors could find applications in electronics and thermally dissipative materials and, at the same time, increase the environmental sustainability of the electronics sector.

## 4. Experimental Section

*Materials*: Recycled printing paper was used as the flexible substrates. GNPs were purchased from XG Sciences (grade M25) and are characterized in previous work.[83] Graphitized CNFs (nominal length between 20 and 200 μm, diameter ≈ 100 nm) were purchased from Sigma-Aldrich (grade PR-25-XT-HHT from Pyrograf Products Inc.) and characterized previously.[30] Flakes of carnauba wax were purchased from Sigma Aldrich and were emulsified in isopropyl alcohol (Sigma Aldrich) following a procedure previously published.[67] For clarity, the preparation of the sample loaded with 30 wt% of nanofillers with respect to CW is described in the following: carnauba wax (0.5 g) was melted in isopropanol (28 ml) using a heat gun as a heat source. The heat gun procedure was stopped when the wax melted (1-2 minutes). The temperature reached by the emulsion is approximately 90°C. GnPs or hybrids of GNP and CNF (0.15 g) were added to the ink. After tip sonication (20 kHz, 40% amplitude, 4 times for 60 s), 4.5 ml of conductive dispersion was spray coated on the paper (7x5 cm$^2$) from a distance of 15-20 cm. A heat gun was used for 30 seconds at approximately 100 °C in order to melt the wax and bind the nanofillers to the paper. For the densified samples, an additional pressing procedure (3 minutes, pressure ranging from 6.25 to 112.5 bar) was performed using Teflon anti attachment foil to prevent the sticking of the paper on the platen of the press.

*Methods*: SEM images of the morphology and of the cross-section of the specimens were taken with a Zeiss Evo50 microscope operating at an acceleration voltage of 10 kV. For cross-sectional SEM analysis, the samples were frozen using liquid nitrogen and fractured.

The electrical features of the paper-based conductors were measured using a source-meter from Keithley (model 2450) in the four-probe configuration. Silver conductive paste (RS pro, product number 186-3600) was coated creating 5 mm wide contacts on the specimens spaced by 5 mm.

Repeated bending cycles were performed on samples to assess mechanical degradation. Samples underwent repeatable and uniform bending cycles while suspended between two supports. One support oscillated horizontally, causing the sample to transition between and a flat and a curved orientation. A pneumatic cylinder (Festo Model ADN-20-50-A-P-A) which was controlled by an electronically switched solenoid valve (Festo Model VUVG Metric M5 5/2) caused the sliding support to move back and forth. During bending cycles, degradation in both the flat and curved orientations was periodically quantified. Four point probe measurements of the surface resistance were taken at regular intervals and normalised to a baseline value, calculated prior to any bending cycles. Four 20 µm thin copper wires were attached to the edges of the sample in a rectangular arrangement. Surface resistance values were obtained by sweeping a DC current (Keithley Model 6221) between two contacts and measuring the resultant voltage across the other two contacts (Keithley Models 2182A). During each sweep, current flow was always along the direction of motion of the oscillating horizontal support.

The folding stability of the electrical features after repeated fold-unfold cycles was recorded measuring the resistance change perpendicular to the folding edge. A weight of ~ 1.5 kg was placed on the fold mark during the folding event.

The EMI shielding effectiveness of the conductors was measured by means of a vector network analyzer (Keysight N5227A) and two WR-90 (8.2-12.4 GHz) waveguides. The transmittance was reported between 8 and 12 GHz.

Abrasion testing was conducted on a Taber Industries 5155 Abraser with CS10 abrading wheels with a 250 g total mass load per wheel. The wheels were refaced with S-11 disks for 50 cycles between each abrasion interval. Tests were conducted 1, 5, 10, 20, 40, 80, 120, 160, 200 and 240 cumulative cycles at 72 rpm.

The thermal diffusivities of the specimens were measured using a custom built setup consisting of a pulsed (1 Hz) tunable laser beam to excite heat waves that propagates periodically into the sample and as using as detector a high resolution infrared camera (FLIR T660). The camera were mounted with an IR micro lens (pixel size 50X50) (Figure 5). Thermal diffusivity is than calculated following the Angstrom method.[84] For each samples eight measurements were performed along the X and Y axis.

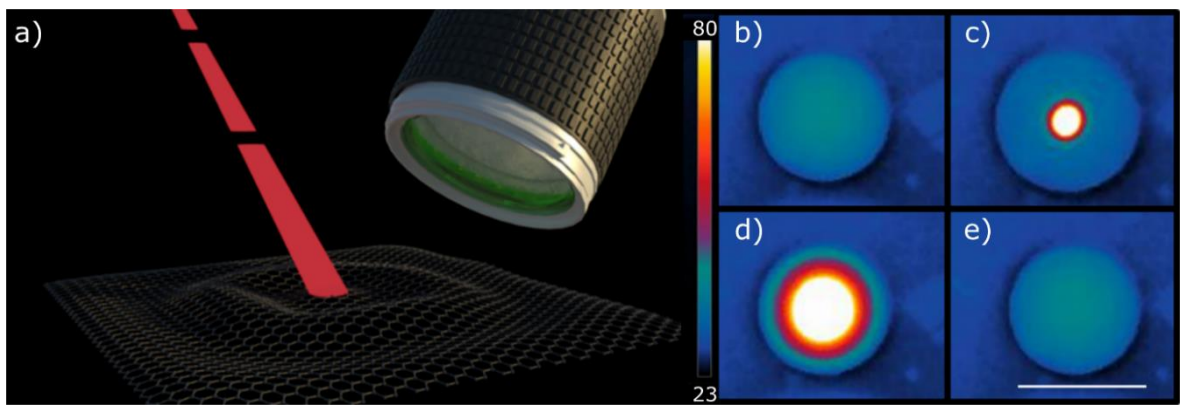

**Figure 5**. a) Schematic representation of the custom-built setup for measuring the thermal diffusivity. The laser is used to produce a periodic heat waves that are recorded by the Infrared camera. b-e) four recorded IR images highlighting the propagation of the heat wave into a sample, scale bar 1 cm.

Thermal gravimetric analysis was carried out on TA Instruments Q5500 TGA on samples of a mass of approximately 5 mg. Samples were heated under a nitrogen atmosphere at a rate of 10 °C/min to 800 °C.

DSC runs were conducted on a TA Instruments Q100 DSC for samples weighing around 1 mg. The samples were ran under the following heat-cool-reheat method: equilibrate at -90 °C, heat at 5 °C/min to 120 °C, cool at -5 °C/min to -90 °C and reheat at 5 °C/min to 120 °C. Values were extracted from the reheat run. An Instron 3365 (pull rate of 1 mm min$^{-1}$) was employed for measuring the stress-strain features of the paper-based conductors. For all the above mentioned measurements and the calculation of the error bars and mean values at least three different samples were employed unless specified differently.

**Acknowledgements**

This project has received funding from the European Union's Horizon 2020 research and innovation programme under grant agreement No 785219.

Received: ((will be filled in by the editorial staff))
Revised: ((will be filled in by the editorial staff))
Published online: ((will be filled in by the editorial staff))

Materials for electronics that are electrically and thermally conductive are rigid, expensive, difficult to be sourced and sometimes toxic. An electrical and thermal conductor that is flexible, lightweight, and environmentally-friendly is manufactured functionalizing paper with a green conductive ink. The ink is made with an alcohol-based emulsion of carnauba wax mixed with nanocarbon materials.

X Wu, P. Steiner, T. Raine, G. Pinter, A. Kretinin, C. Kocabas, M. Bissett, P. Cataldi

**Hybrid Graphene/Carbon Nanofiber Wax Emulsion for Paper-based Electronics and Thermal Management**

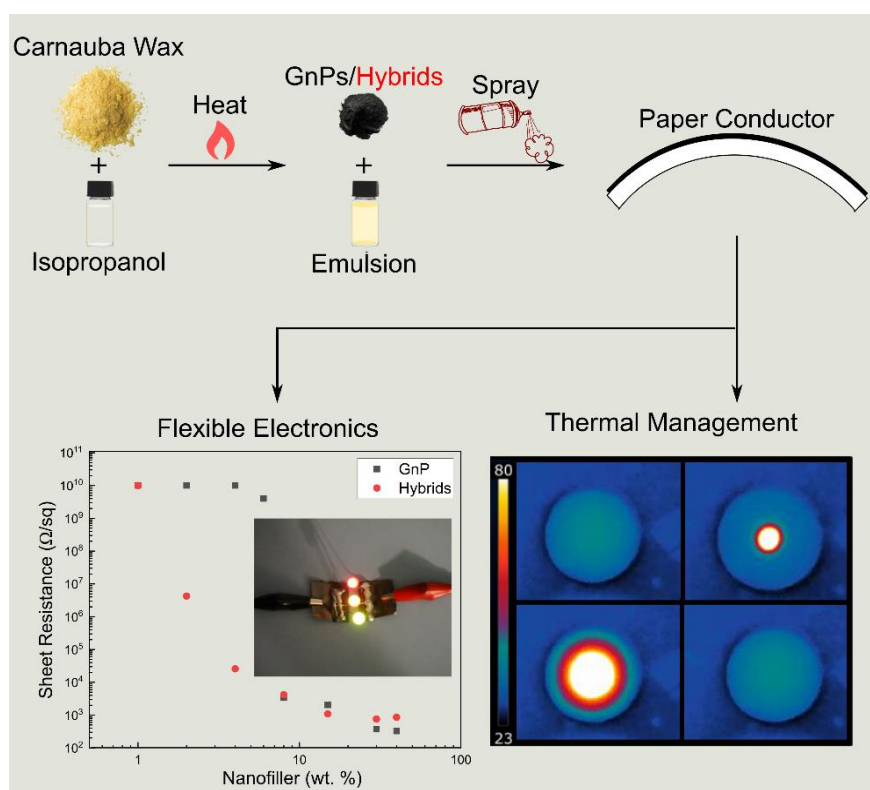

# Hybrid Graphene/Carbon Nanofiber Wax Emulsion for Paper-based Electronics and Thermal Management

*Xinhui Wu, Pietro Steiner, Thomas Raine, Gergo Pinter, Andrey Kretinin, Coskun Kocabas, Mark Bissett\*, Pietro Cataldi\**

X Wu, P. Steiner, T. Raine, G. Pinter, A. Kretinin, C. Kocabas, M. Bissett, P. Cataldi

Department of Materials, National Graphene Institute, University of Manchester, Oxford Road, Manchester, M13 9PL UK
E-mail: pietro.cataldi@manchester.ac.uk; mark.bissett@manchester.ac.uk

Differential Scanning Calorimetry

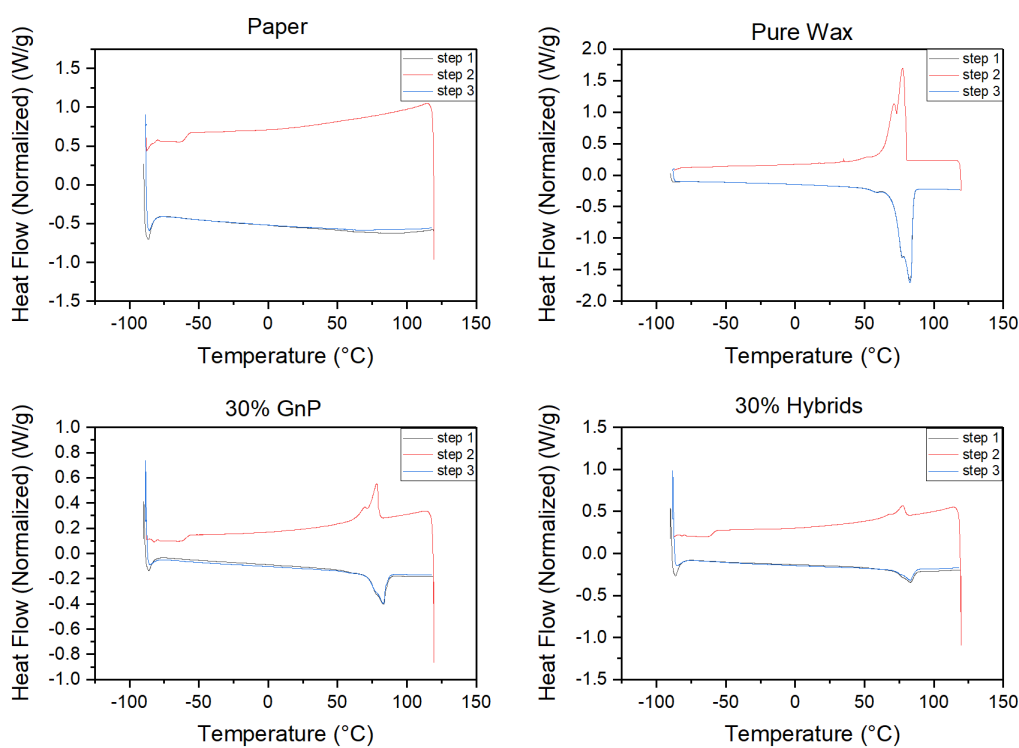

Figure S 1: Differential scanning calorimetry of a) the paper substrate, b) the pure carnauba wax, c) the sample with 30 wt.% of GNPs and d) the specimen with 30 wt.% of hybrid nanofillers.

Carnauba Wax Particle Size Distribution

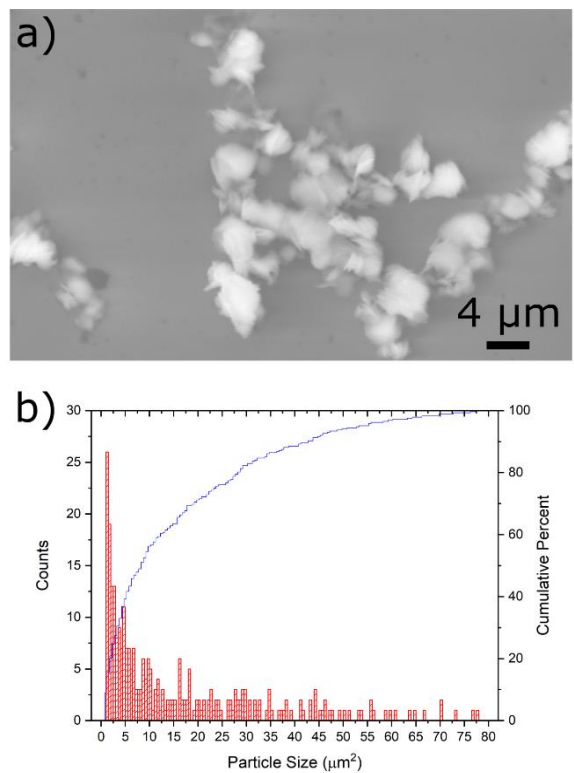

Figure S 2: a) SEM representative image of the CW micro-particles obtained after that the emulsion cooled down. A highly diluted solution (0.02 mg/ml) was drop casted on a silicon substrate and, after solvent evaporation, the micro-particles pictures were acquired. b) Statistical analysis of the particle size distribution obtained counting approximately 290 CW micro-flakes.

Thermogravimetric Analysis

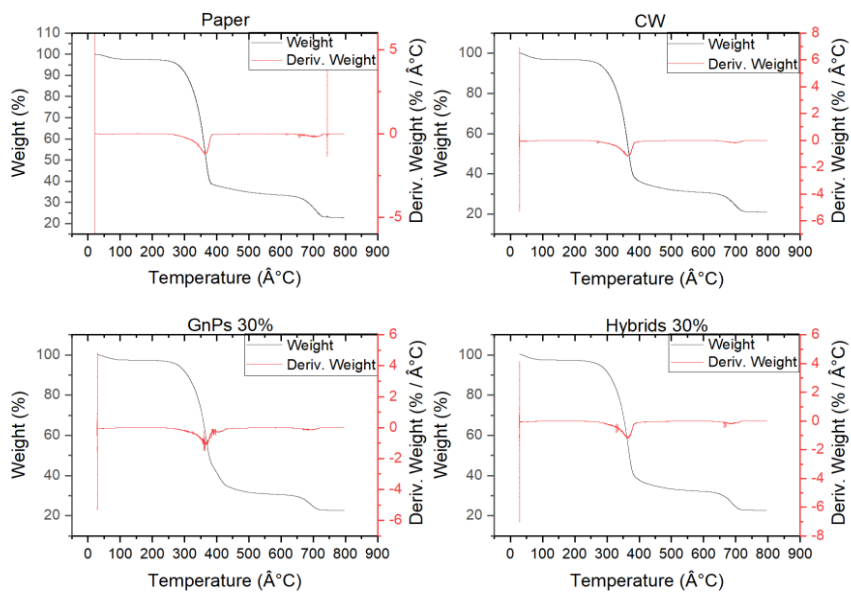

Figure S 3: Thermogravimetric analysis of the samples. Weight loss and its derivative with respect to temperature are shown.

Linear Fit and Critical Exponent Calculation

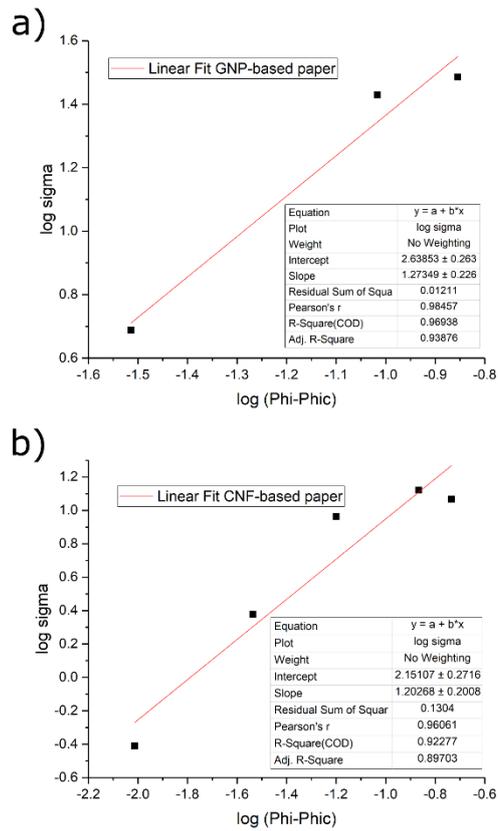

Figure S 4: Linear fit to extract the critical exponent from the formula $\sigma = \sigma_0 (\phi - \phi_c)^t$. a) and b) are the linear fit related to the GNP- and hybrid-based material, respectively. Both gives a universal critical exponent of approx. 1.2-1.3.

Electrical Features Bending Stability

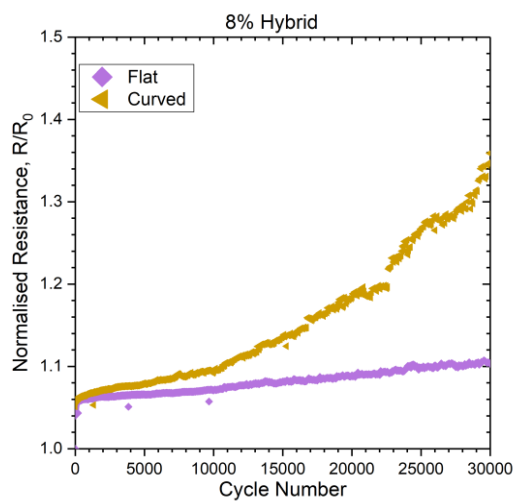

Figure S 5: Bending stability of the 8 wt.% hybrid loaded sample. The GnP-based sample was exhibiting worse results.

Mechanical Properties

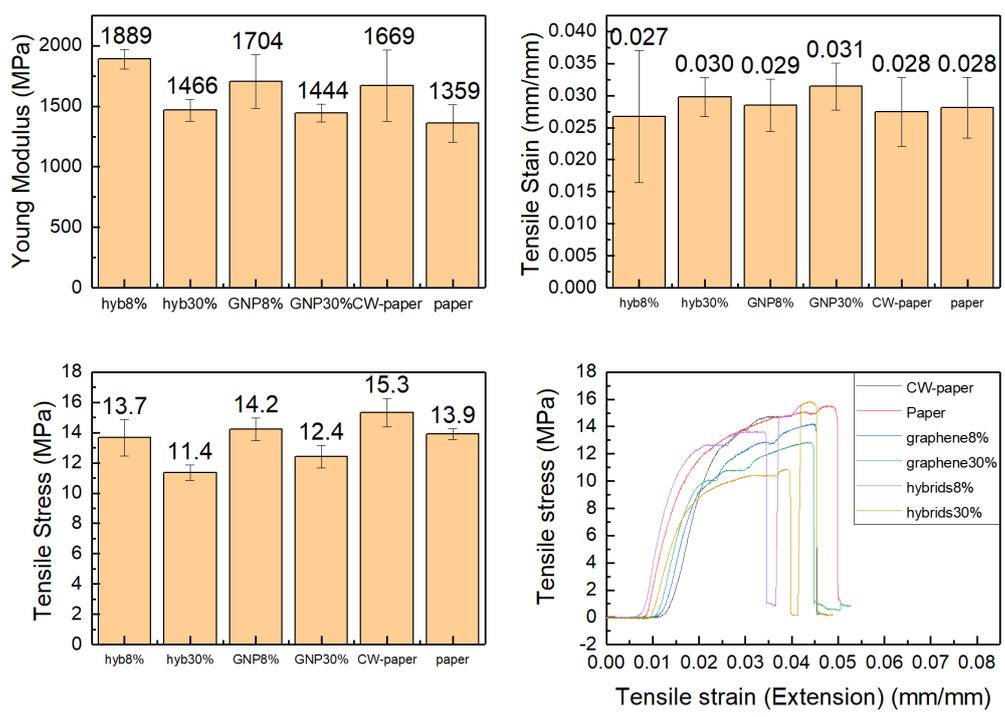

Figure S 6: a), b), c) Young modulus, tensile strain and tensile stress extracted from the d) stress strain curves, respectively.

SEM Morphology of the Pressed Sample

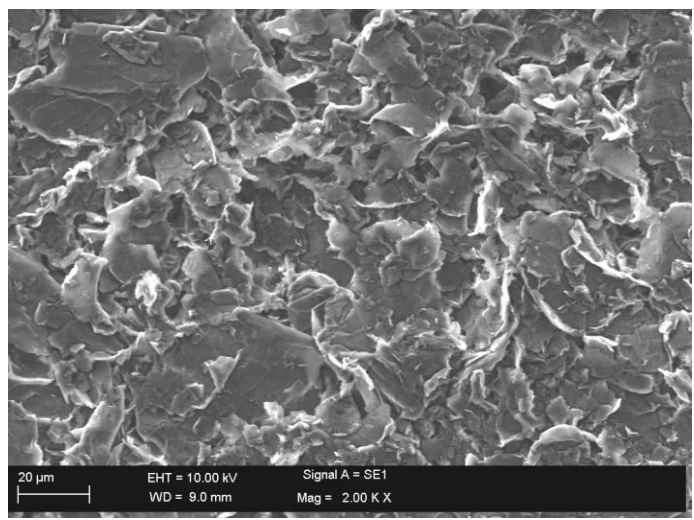

Figure S 7: SEM image of the 30 wt.% GNP-based coating after press at 112.5 bar.

SEM Cross Section of the Pressed Sample

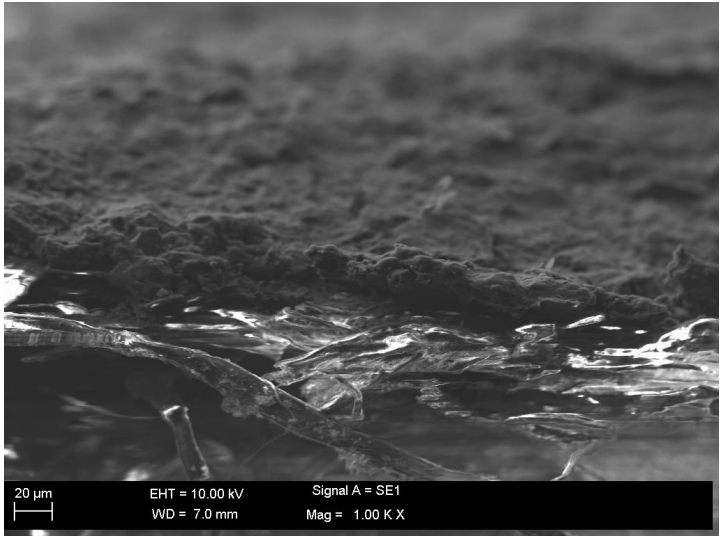

Figure S 8: SEM cross sectional image of the 30 wt.% Hyb-based coating after press at 112.5 bar.

Bending Stability of the Electrical Properties after Press

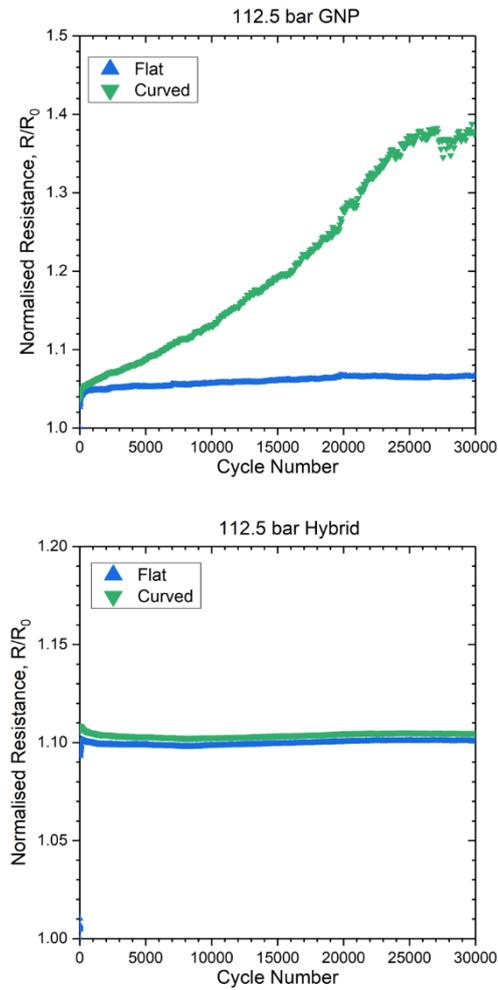

Figure S 9: Bending Stability of the electrical resistance of the sample after press at 112.5 bar. The top image refers to the 30 wt.% GnP-based sample, the bottom to the hybrid one with the same concentration of nanofillers.

SEM after Abrasion Cycles

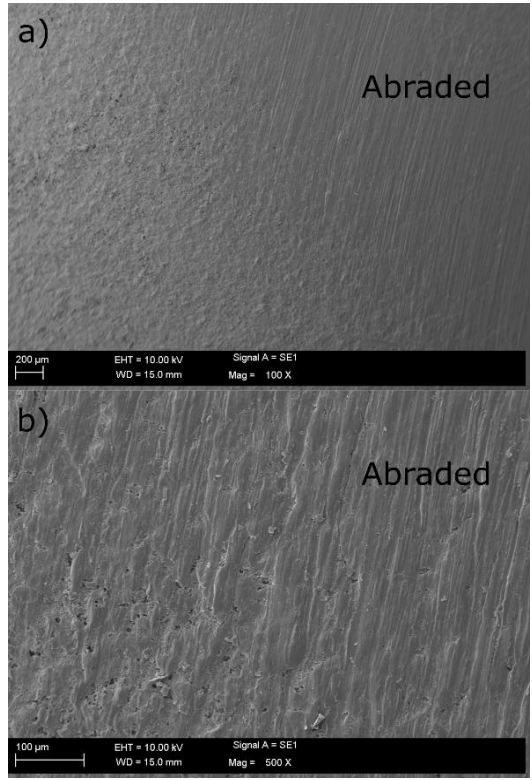

Figure S 10: SEM of the morphology of the 30 wt.% GNP pressed sample after 200 abrasion cycles.

EMI Shielding of the Paper-based Conductor

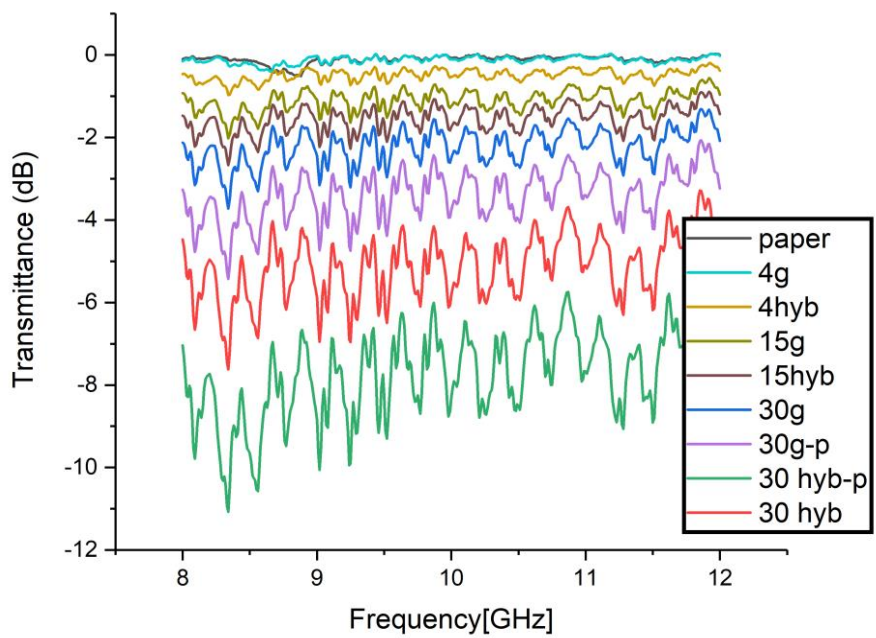

Figure S 11: Emi shielding performance of the conductive paper between 8 and 12 GHz. The number in the label express the concentration of the nanofiller with respect to the weight of the carnauba wax.